\documentclass[letterpaper]{article}
\usepackage{isea}
\usepackage[pdftex]{graphicx}
\usepackage{times}
\usepackage{helvet}
\usepackage{courier}
\usepackage{cuted}
\usepackage{graphicx}
\usepackage{newfloat}
\usepackage{caption}
\usepackage{xurl}
\usepackage[colorlinks=true, urlcolor=black, citecolor=black, linkcolor=black]{hyperref}

\usepackage[numbers]{natbib}
\title{The Dream Within Huang Long Cave: AI-Driven Interactive Narrative for Family Storytelling and Emotional Reflection}
\author{Jiayang Huang\textsuperscript{1}, Lingjie Li\textsuperscript{2},Kang Zhang\textsuperscript {3}, David Yip\textsuperscript{4}\\
\textsuperscript{1} Hong Kong University of Science and Technology (Guangzhou), China. jhuang130@connect.hkust-gz.edu.cn\\
\textsuperscript{2} Sungkyunkwan University, South Korea. lijingjie@skku.edu\\
\textsuperscript{3} Hong Kong University of Science and Technology (Guangzhou), China. kzhangcma@hkust-gz.edu.cn\\
\textsuperscript{4} Hong Kong University of Science and Technology (Guangzhou), China. daveyip@hkust-gz.edu.cn\\
\newline
}

\begin{document} 
\maketitle


\begin{abstract}
This paper introduces the art project \emph {The Dream Within Huang Long Cave,} an AI-driven interactive and immersive narrative experience. The project offers new insights into AI technology, artistic practice, and psychoanalysis. Inspired by actual geographical landscapes and familial archetypes, the work combines psychoanalytic theory and computational technology, providing an artistic response to the concept of “the non-existence of the Big Other.” The narrative is driven by a combination of a large language model (LLM) and a realistic digital character, forming a virtual agent named YELL. Through dialogue and exploration within a cave automatic virtual environment (CAVE), the audience is invited to unravel the language puzzles presented by YELL and help him overcome his life challenges. YELL is a fictional embodiment of the “Big Other,” modeled after the artist's real father. Through a cross-temporal interaction with this digital father, the project seeks to deconstruct complex familial relationships. By demonstrating “the non-existence of the Big Other,” we aim to underscore the authenticity of interpersonal emotions, positioning art as a bridge for emotional connection and understanding within family dynamics.
\end{abstract}

\keywords{Keywords}
Interactive Narrative, AI-driven Family Storytelling, Immersive Art Installation, Psychoanalysis, Art Practice.

\begin{figure}[h]
\includegraphics[width=3.31in]{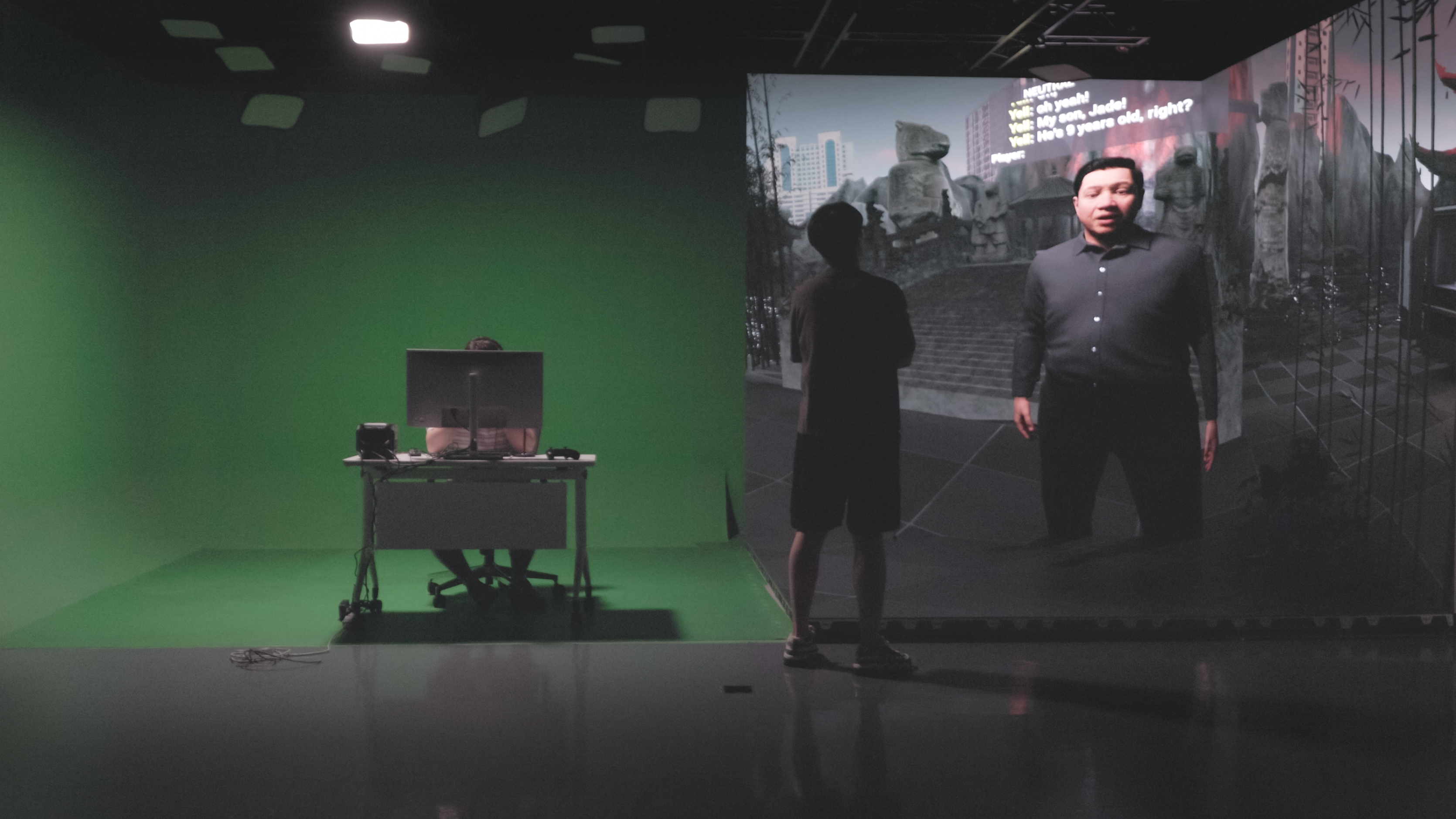}
\caption{The hybrid immersive environment of art installation combined with a green screen and CAVE system.
\url{https://vimeo.com/1023900209}} 
\end{figure}

\section{Introduction}
\subsection{The crisis in family relationships}
With globalization, material life has become more abundant, yet mental and psychological needs remain unmet. According to the World Health Organization (WHO) and the Organisation for Economic Co-operation and Development (OECD), mental health issues have significantly increased, with depression affecting around 300 million people globally in 2017, particularly in developed countries \cite{world2017depression} \cite{oecd2017}. This growing trend is not only driven by increasing social pressures but also deeply influenced by underlying family dynamics.

Family relationships play a crucial role in an individual's psychological health. Freud's Oedipus complex suggests that emotional attachments formed in childhood shape self-identity, gender roles, and social behavior \cite{freud1924loss}. Similarly, Bowlby's attachment theory highlights how early relationships with caregivers impact personality development, with insecure attachments leading to anxiety in adulthood \cite{bowlby1982attachment}. Bronfenbrenner also emphasized that family dynamics are central to socialization and mental health \cite{bronfenbrenner1979ecology}. These theories collectively show that family stability and healthy relationships are vital to an individual's psychological development.

However, contemporary families face growing communication barriers, especially between generations. A 2020 American Psychological Association (APA) report revealed that these issues often arise from generational value differences, cultural conflicts, and the fast pace of modern life \cite{american2020stress}. In Eastern cultures, Confucian norms emphasize the father's authority, which, while maintaining order, exacerbates emotional distance and limits children's self-expression, leading to a lack of identity and support \cite{slote1998confucianism}. These unresolved barriers, characterized by distant or emotionally detached family relationships, often contribute to depression, anxiety, and other mental health issues in adulthood  \cite{repetti2002risky}.

\subsection{From the Name-of-the-Father to the Big Other}
The structure of familial authority can be understood through Lacan’s concept of the Name-of-the-Father, which signifies paternal law and symbolic authority. This idea builds on Freud’s discussion of the “primordial father” in \emph {Totem and Taboo(1913)}, where the father figure represents absolute power and prohibition. Lacan extends this notion, arguing that the Name-of-the-Father does not refer to an actual father but rather a symbolic function that governs desire and social norms \cite{miller2018four}. Over time, this authority became absorbed into a broader symbolic system—what Lacan terms the Big Other. The Big Other encompasses language, law, and cultural structures that shape individual subjectivity and mediate interpersonal relationships \cite{lacan2013psychoses}.

However, Lacan later proposed that “the Big Other does not exist” (Il n’y a pas de Grand Autre), emphasizing the instability of the symbolic order \cite{lacan2001ecrits}. While the Big Other appears to function as an external authority that provides meaning and coherence, it is ultimately an illusion sustained by collective belief. This insight is crucial for understanding the contradictions within familial relationships—rigid symbolic structures often conceal inherent gaps and inconsistencies. By exposing these fractures, it becomes possible to challenge dominant narratives of paternal authority and reconstruct more authentic emotional connections. 

\subsection{Psychoanalytic art practice}
This study explores Lacan's concept of the “non-existence of the Big Othe” through the interactive narrative artwork, \emph{The Dream Within Huang Long Cave}. Using psychoanalytic principles, artistic contextualization, and interactive design, the work deconstructs the authority of the Name-of-the-Father and the Big Other, bringing genuine memories to the forefront. Through AI-driven family storytelling, audiences engage in dialogues with a fictional father figure, revealing linguistic inconsistencies that expose the illusionary nature of the symbolic order. Ultimately, this psychoanalytically artistic practice encourages reflection on family dynamics, societal authority, and individual subjectivity by fostering a space for emotional connection and critical insight.

\section{Background}
\subsection{The otherness in psychoanalysis}
The concept of the “other" has been a fundamental concern in Western philosophy, evolving from early dialectical and existential thought to its central role in psychoanalysis. While classical philosophy explored self-consciousness through relational dynamics, it was Lacan who systematically redefined the concept within the realm of psychoanalysis \cite{lacan2013psychoses}. Building on Freud’s differentiation of the id, ego, and superego, Lacan distinguished between the little other—a reflection of the ego in the Imaginary order—and the Big Other, which represents the embodiment of the Symbolic that governs language, law, and social structures \cite{evans2006introductory}.

Lacan argues that subjectivity is fundamentally mediated by the Big Other, as “the unconscious is the discourse of the Other” \cite{lacan2001ecrits}. However, this symbolic authority is not a fixed entity but an illusion sustained by discourse, revealing structural gaps and inconsistencies. This notion is further developed by Žižek, who analyzes how the Big Other manifests in popular culture, shaping desires and ideological narratives through films, advertisements, and political discourse \cite{vzivzek2006pervert}. These cultural representations simultaneously reinforce and expose the instability of symbolic authority, making them critical sites for its deconstruction.


\subsection{AI as the contemporary Big Other}
According to Lacan's well-known assertion that “Desire is the desire of the Other,” individual desire is indirectly constructed within the symbolic order through the desire of others \cite{lacan2011seminar}. Desire stems from an inherent lack (manque) within the subject, which implies that the subject tries to compensate for this lack by conforming to the expectations of the Big Other. In this sense, artificial general intelligence (AGI) represents a human fantasy projected outward to compensate for our shortcomings. Millar’s concept of the “Sexbot” illustrates this notion, with AI-based humanoid figures like Ava in \emph{Ex Machina (2014)} and Joi in \emph{Blade Runner 2049 (2017)} representing both the fulfillment and impossibility of human desires \cite{millar2021psychoanalysis}. Similarly, Luca's work emphasizes that AI is not merely an engineering construct but a technological reflection of the unconscious, reinforcing Lacan’s theories of the mirror stage and the Oedipal complex \cite{possati2020algorithmic}. Mar and Anne's study highlights how AI's “gaze” has become central to surveillance systems, describing it as a “demonic gaze” that reflects Sartre's “gaze of the Other” \cite{heimann2023stainless}.

Moreover, the emergence of large language models (LLMs) has positioned AI as a symbolic repository of knowledge—an encyclopedic Big Other. Through training on vast corpora, LLMs have become experts, possessing extensive knowledge reserves and linguistic abilities. In this paper, the authors evaluate the capabilities and limitations of LLMs, highlighting their accuracy and expertise in scientific exploration\cite{adesso2023towards}. Additionally, another paper demonstrated that GPT-4 exhibits high accuracy when answering questions related to surgical knowledge, proving that LLMs can perform at an expert level in specific domains \cite{beaulieu2024evaluating}. This work shows LLMs' applications across different fields, including finance, engineering, and law, demonstrating their significant capability in cross-disciplinary problem-solving and knowledge integration. These characteristics point to LLMs being the absolute authority of knowledge, embodying the omniscience of the Big Other \cite{hadi2024large}.

\subsection{Art as a means of revelation}
From a psychoanalytic perspective, Lacan viewed art as a mechanism for grasping the unspeakable Real, distinguishing it from mere illusion \cite{levine2008lacan}. Through sublimation, art transforms unconscious desires into symbolic expression. For example, da Vinci’s \emph{Mona Lisa  (1503)} sublimes the gaze of desire, while Dalí’s \emph{The Persistence of Memory (1931)} distorts time as a metaphor for the unconscious. Art operates as a “rope” stabilizing us over the void, revealing fragments of the Real beneath the symbolic order. Expanding on this, Žižek argues that cinema functions as the “ultimate perverted art” because it dictates how we desire rather than merely fulfilling desire \cite{zizek1992looking}. Hitchcock’s films, such as \emph{Vertigo (1958)} and \emph{Psycho (1960)}, destabilize reality by exposing the contradictions between subjective fantasy and external symbolic structures.

Building on this, new media art extends the role of art in disrupting dominant symbolic frameworks. Joline and Jon propose the concept of “art as an antibody,” arguing that art critically intervenes in and subverts hidden ideological structures \cite{blais2006art}. Digital and algorithmic art, as seen in early internet-based works, exposes societal norms through irony and disruption. Furthermore, video games, with their interactive and immersive nature, offer a novel mechanism for engaging with the Real. Jean and Sarah suggest that video games preserve not only visual and spatial aspects of an artwork but also the emotional responses they evoke \cite{bridge2009preserving}. This capacity to integrate audience interaction allows new media art to challenge the symbolic order dynamically, making the experience of the Real more tangible.

\section{Related Works}
This section classifies and analyzes new media artworks that inspired our project, focusing on three core characteristics: otherness, AI’s technicality, and emotionality in video game art.

David OReilly's \emph {Everything (2017)} exemplifies otherness by allowing players to embody perspectives at multiple levels, from molecular to galactic, invoking the wonder of animism and prompting reflections on the self beyond the physical body \cite{Graf_2017}. In \emph {Redundant Assembly (2015)}, Rafael's installation acts as a “technological mirror,” showing the audience an alienated version of themselves \cite{Lozano-Hemmer_2015}. In \emph {Lights! Dance! Freeze! (2023)}, Theodoros matches viewer movements with cinematic characters using search algorithms, reflecting their physical posture through film frames \cite{papatheodorou2023lights}.

For works focused on AI's technicality, Yuqian's \emph {1001 Nights (2022)} positions AI as a language repository, where players use keywords to activate the AI and uncover stories \cite{sun2023language}. Weidi's \emph {ReCollection (2023)} employs AI-generated technology to visualize memory, describing AI as a non-human narrative agent that bridges language, imagination, and memory \cite{zhang2024recollection}. Similarly, Purav's \emph {Ghost in the Machine (2023)} uses multimodal AI to generate audiovisual responses to participants, revealing AI as an “illusion of intelligence” \cite{bhardwaj2023ghost}.

In the emotional dimension of video game art, Kallina's \emph {ManicVR (2018)} creates a virtual experience based on personal family history, evoking empathy for siblings' mental struggles. Similarly, Julian's \emph {Promesa (2020)} and Marion's \emph {EMPEROR (2024)} draw from family stories. In \emph {Promesa}, surreal hometown scenes convey nostalgia through “absence of presence”. In \emph {EMPEROR}, the story of a father with aphasia unfolds with monochrome visuals, revealing fragile yet warm relationships between individuals.


\section{Research Question and Methodology }
This study examines the role of AI-driven artistic practice in engaging with Lacan’s proposition that "the Big Other does not exist." We ask: What happens when AI is positioned as a tangible representation of the Big Other? Can an AI-driven character, modeled after a paternal figure, reveal the instability of symbolic authority and disrupt rigid familial structures through interactive storytelling? We hypothesize that by embodying the symbolic father, AI can serve as both a subject of engagement and a site of deconstruction, allowing audiences to question established familial roles and experience emotional shifts through gameplay. Why do interactive narratives provide a more effective medium for challenging symbolic structures than traditional storytelling? How does engaging in dialogue with an AI alter perceptions of paternal authority and emotional connection? These questions help define the potential of AI-driven storytelling as a method for deconstructing familial authority and reconstructing emotional narratives.

\begin{figure}[h]
\includegraphics[width=3.35in]{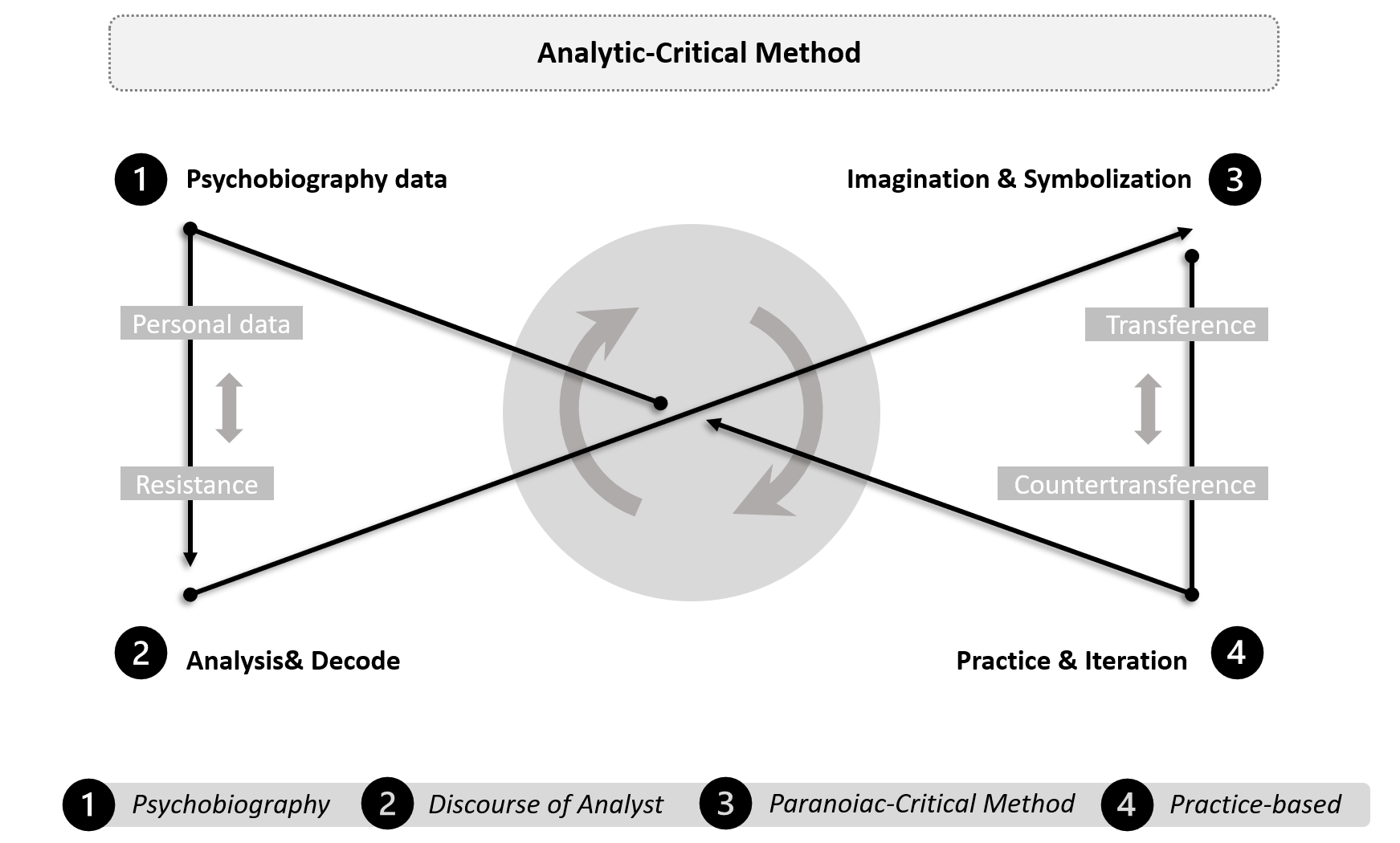}
\caption{The diagram of the Analytic-Critical Method, integrating four research methods.} 
\end{figure}

To address these research questions, this study introduces the Analytic-Critical Method (Fig.2), a four-step framework designed to construct the AI-driven Father character and interactive experience. First, we employed psychobiography to collect personal historical data and explore unconscious desires \cite{freud1964leonardo}. Through interviews, monologues, and archival materials, we documented the artist's father's memories, reconstructing his early life and emotional landscape. This dataset provided the foundation for AI-Father’s persona.

Second, inspired by Lacan’s discourse of the analyst, we analyzed and decoded the father’s language to uncover latent unconscious structures \cite{lacan2007seminar}. The artist assumed the analyst’s role, examining the father’s speech for patterns of repression, desire, and resistance. Through this process, unconscious elements—such as fears embedded in nightmares, aspirations in fantasies, and slips of speech—were extracted and symbolically encoded into the AI character. These elements aligned with Lacan’s objet petit a, the unattainable remainder of desire, shaping the emotional complexity of AI-Father \cite{miller2018four}. 

Third, inspired by Dalí’s paranoiac-critical method, the artist engaged in a reverse perspective, imagining himself as the father \cite{finkelstein1975dali}. This process allowed for a deeper understanding of the character’s psychological motivations, enabling the artist to design AI-Father’s behaviors and reactions in emotionally challenging scenarios. The AI-Father was constructed to reflect both the artist’s projections and the father’s unconscious struggles, making the interaction more dramatic and psychologically resonant. However, during this process, both the father and the artist encountered resistance—manifesting as transference—where the artist’s own “lack” was projected onto the AI character \cite{lacan2012seminar}.

Finally, the project incorporated practice-based art research, following iterative cycles of planning, execution, and critical reflection \cite{candy2021practice}. The AI-Father’s dialogue model, visual representation, and interactive mechanics were refined through audience engagement and feedback loops. This iterative design allowed the work to evolve dynamically, ensuring that the AI-Father’s interactions effectively embodied both the symbolic authority and its deconstruction. The process also facilitated counter-transference, enabling the artist to maintain a critical distance while navigating the emotional complexities of familial reconstruction \cite{lacan2012seminar}. 

Through these four methodological steps, the study bridges psychoanalysis, AI technology, and interactive storytelling, ultimately crafting an artistic experience that challenges symbolic structures and reconstructs emotional connections.

\section{Implementation}
\subsection{Unconscious subject design}
The objective of \emph{The Dream Within Huang Long Cave} is to create an interactive family narrative centered on the father figure. Rather than adopting a conventional character design approach, this study employs a psychoanalytic framework to construct the father as an “unconscious subject,” shaped by his desires, lacks, and evolving self-perception \cite{lacan1988ego}. Through the Analytic-Critical Method, we identified a defining formative element in the father’s unconscious: his longing for an absent father. This absence, which created a deep emotional void, became a crucial reference point in constructing the AI-father’s character.

Recognizing that an individual’s desires and unconscious motivations are not static but evolve over time, we designed three versions of the AI-father, representing his life at ages 11, 24, and 36. The selection of these three stages was not arbitrary but was informed by both psychological and biographical justifications. Each stage corresponds to a critical period of identity transformation, where shifts in familial roles and external circumstances profoundly influenced the father’s sense of self. At age 11, the father underwent a pivotal environmental shift after his family relocated. Previously introverted and obedient, he became more rebellious, reflecting the impact of social adaptation on personality formation. At age 24, he entered adulthood and took on familial responsibilities following his marriage, facing the tension between personal ambition and family obligation. By age 36, the father experienced the loss of his own father, a fundamental psychoanalytic shift in subjectivity. This event marked a forced transition from the position of the son to that of the father. In Lacanian terms, he moved from being subjected to the Name-of-the-Father to embodying it himself, taking on the very role of authority that had once shaped—and constrained—him. This dynamic underscores the central tension of the project—how one becomes the figure of authority they once opposed, perpetuating the cycle of symbolic order even as they struggle to break free from it.

Following this analytical and symbolic process, the fictional father character was named “YELL.” The act of naming serves a dual function within the psychoanalytic framework. First, it symbolically reverses Lacan’s concept of the “Name-of-the-Father”, positioning the artist, as the child, in the role of the one who assigns meaning to the paternal figure. Second, the name embodies the AI-father’s emotional core, representing a subject in a perpetual state of calling out (“yelling”) for recognition, connection, and love. Furthermore, the phonetic resemblance between “YELL” and the father’s real name reinforces the project’s psychoanalytic inquiry, intertwining personal memory, artistic reimagination, and symbolic representation.

\begin{figure}[h]
\includegraphics[width=3.31in]{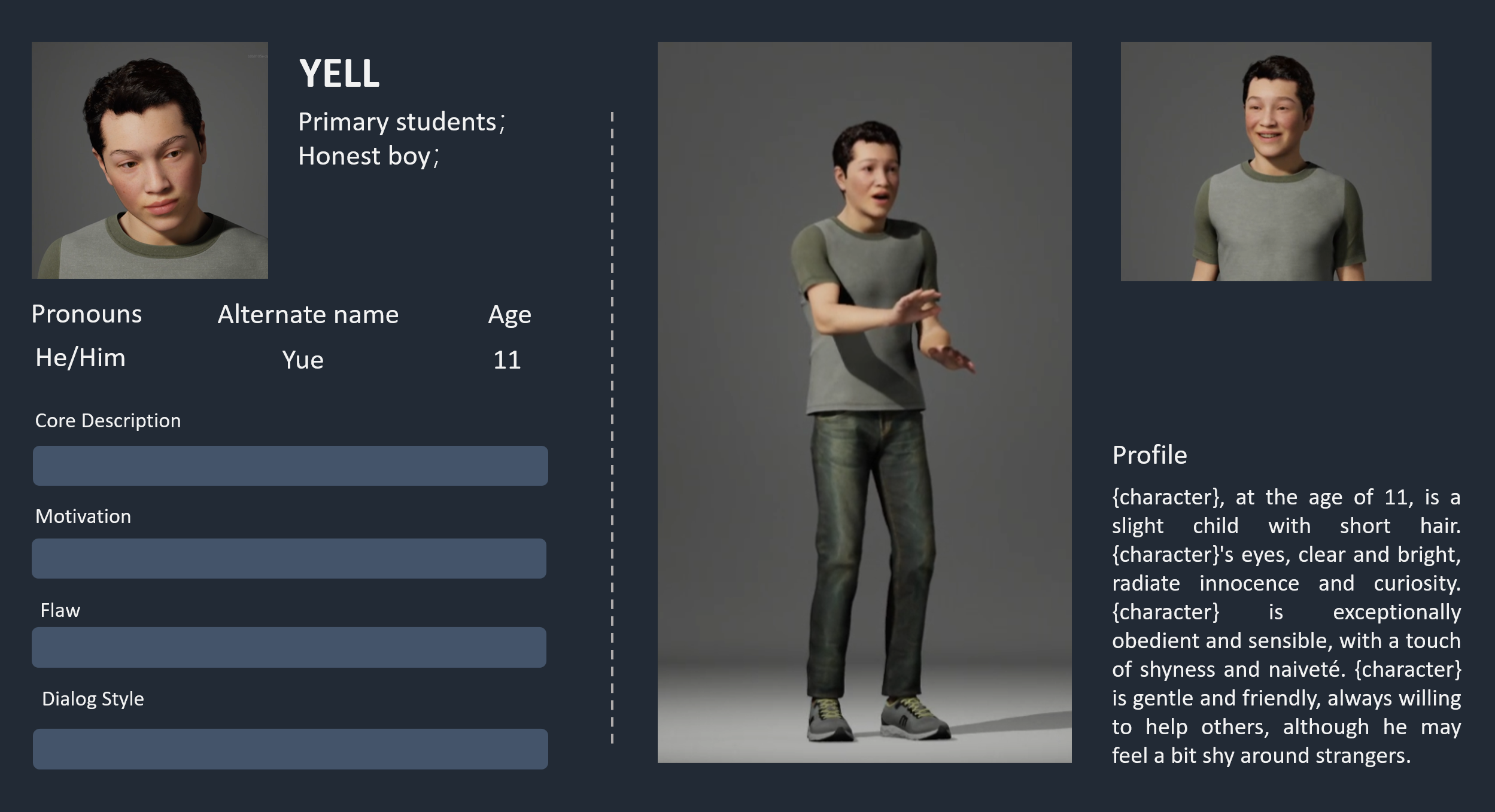}
\caption{The character configuration interface for YELL (age 11).} 
\end{figure}

\subsection{AI-Driven digital father}
With the father’s unconscious structure established, the project required an intelligent mediator to embody YELL, enabling both storytelling and audience interaction. To achieve this, we integrated two key technologies: a GPT-based LLM as YELL's cognitive engine and Metahuman for his visual representation (Fig. 3). Specifically, the Inworld AI plugin \footnote{\url{https://inworld.ai}} was employed to develop YELL’s backstory, incorporating collected personal data, including motivations, personality traits, psychological flaws, and conversational style. This enriched dataset allowed the GPT-based model to generate responses that were more contextually adaptive and emotionally nuanced.

For YELL’s visual representation, we used Metahuman \footnote{\url{https://www.unrealengine.com/metahuman}} to create realistic, age-specific avatars, corresponding to the three life stages identified in our analysis. Rather than digitally replicating the real father, we referenced archival photographs to design YELL’s facial features and expressions at each stage of life. This integration of LLM-driven interaction and Metahuman’s photorealistic rendering was realized within Unreal Engine, providing a symbolic container for YELL as a manifestation of the Big Other—transformed from a projected fantasy into a tangible, interactive presence.

\subsection{Interactive story design}

This section outlines the narrative structure of \emph {The Dream Within Huang Long Cave} and the interactive mechanisms that shape YELL’s transformation across three critical life stages. The story unfolds through dialogue-based interaction, environmental exploration, and symbolic intervention, allowing players to actively engage in YELL’s emotional and psychological development.
\begin{figure}[h]
\includegraphics[width=3.31in]{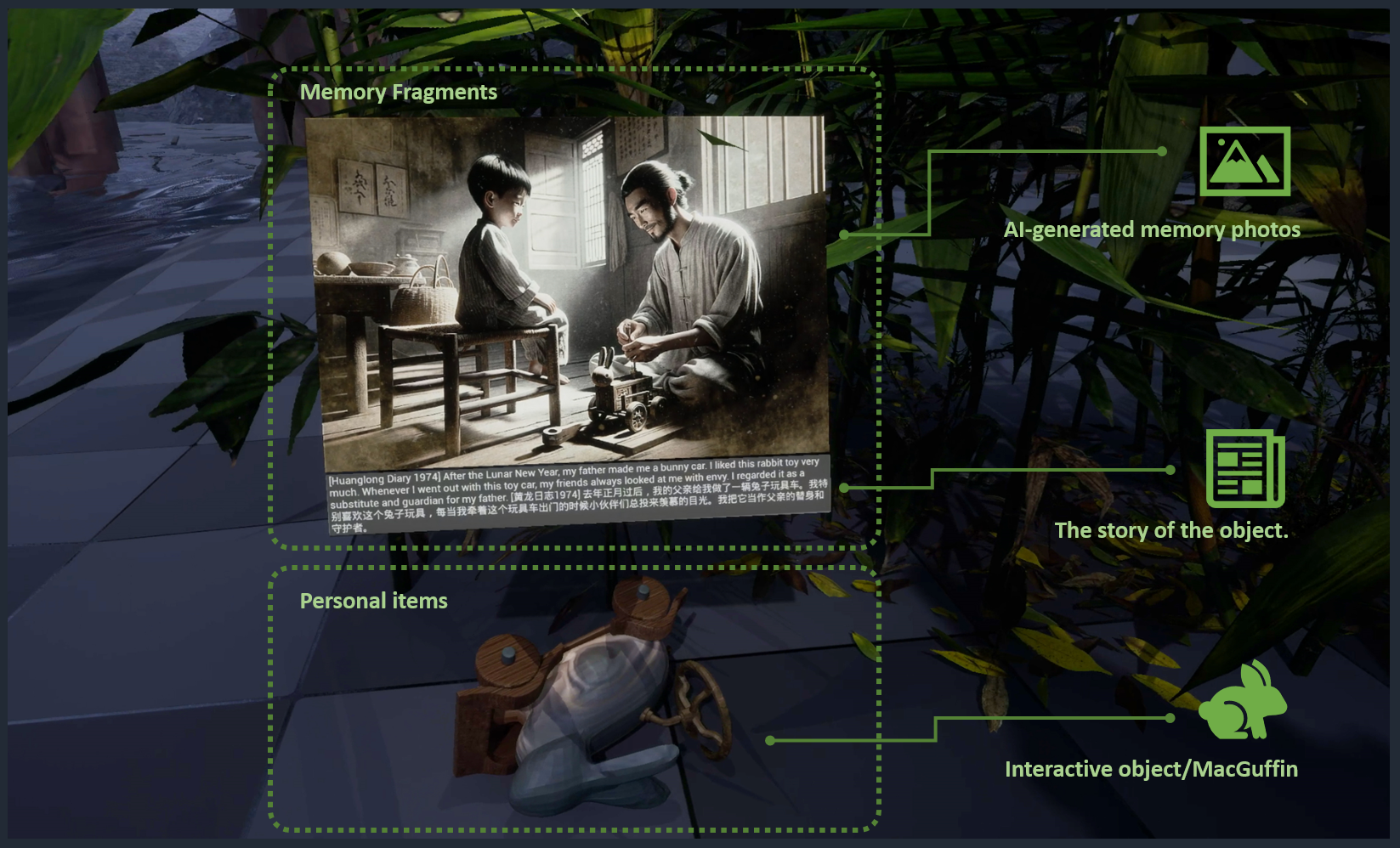}
\caption{Game screenshot illustrating the fragmented memory story and the lost rabbit toy as a MacGuffin.} 
\end{figure}

The first stage introduces 11-year-old YELL, whose primary desire is to find his father. However, this desire is obstructed by internalized fear of paternal authority, causing hesitation and emotional conflict. The player’s role is to interact with YELL, uncovering his concerns and encouraging him to take action. To facilitate this transformation, we introduce Hitchcock’s MacGuffin as a symbolic trigger for YELL’s psychological shift \cite{hitchcock2014hitchcock}. During interviews, the father recalled a rabbit toy car, handmade by his own father, which he had cherished as a treasure (Fig. 4). In the game, players must discover the hidden backstory of this object and share it with YELL, allowing his sense of lack to be temporarily filled, giving him the courage to move forward in his search for his father. This pivotal moment enables both YELL and the player to progress to the next stage.

In the second stage, 24-year-old YELL appears as a young man with ambitious aspirations, envisioning himself as an entrepreneur with grand business plans (Fig. 5a). However, his ideas remain vague, impractical, and disconnected from reality, functioning as a defense mechanism to avoid real-world responsibility. The player’s task is to explore the environment, uncovering memory fragments that reveal the inconsistency in YELL’s claims. Through this process, YELL’s self-deception is exposed, forcing him to confront the disparity between his idealized ambitions and actual capabilities. 

By the third stage, 36-year-old YELL has become consumed by work and social obligations, inadvertently neglecting his family (Fig. 5a). In this chapter, he experiences a nightmare about forgetting something crucial, which the player must decode through exploring and reasoning. As the game progresses, the player pieces together contextual clues, eventually uncovering that YELL has forgotten to pick up his child from school. The player’s intervention not only restores YELL’s awareness of his role as a father but also challenges his internalized symbolic obligations, urging him to re-evaluate his emotional priorities.

Structurally, the game forms a narrative loop, where YELL’s initial search for his father ultimately leads him to recognize his own role as a father. This cyclical transformation embodies Lacan’s concept of “extimacy”, illustrating how external symbolic authority shapes internal desires and how these structures can be disrupted through emotional engagement \cite{lacan2011seminar02}. The interactive format encourages players to actively participate in dismantling rigid paternal ideals, reinforcing the project’s central inquiry into emotional authenticity beyond symbolic constraints.

\begin{figure}[h]
\includegraphics[width=3.31in]{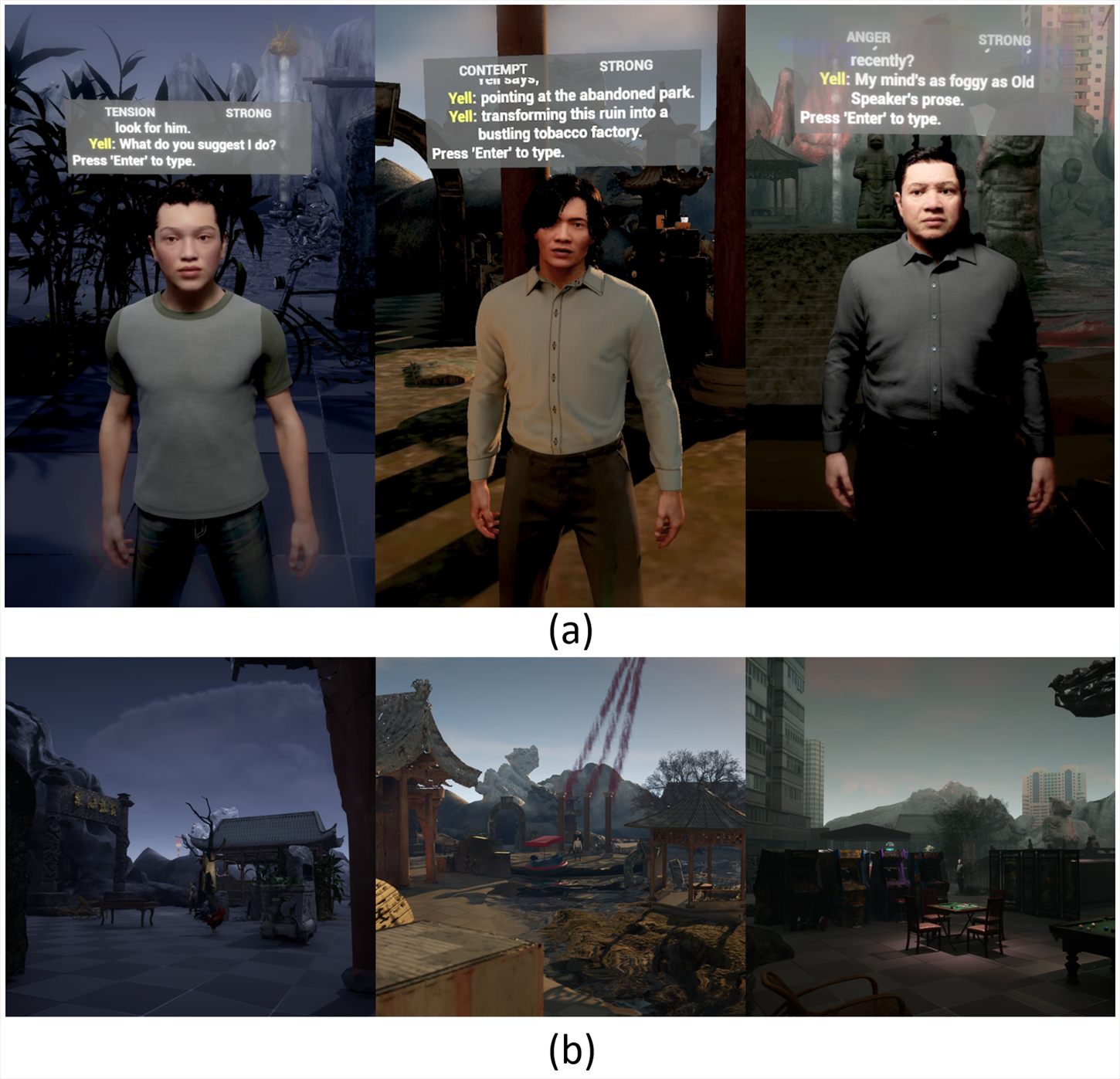}
\caption{(a) YELL character at three different life stages.
(b)Visual environments of the three caves.} 
\end{figure}

\subsection{Scene design}
The story takes place in a cave environment that merges real and imaginary elements, inspired by the actual site of Huang Long Cave near the artist's home. The cave’s design draws on symbolic traditions, from Plato's \emph { Allegory of the Cave} to Kafka's \emph {The Burrow}. The cave always symbolizes mysterious and unconscious realms. In this work, it not only serves as a bridge connecting shared memories between the artist and his father but also represents the symbolic order and the role of the Big Other. Each cave represents a distinct historical and social backdrop (Fig. 5b): the rural simplicity of the 1970s (boy YELL), the economic transformation of the 1980s (young adult YELL), and the urban materialism of the 1990s (adult YELL). These environmental transitions mirror YELL’s evolving subjectivity, reflecting the external pressures that shape his internal struggles.


Throughout the game, various personal items serve as “memory fragments” that aid players in solving puzzles and understanding YELL’s growth journey. These items, including a rabbit toy, match gun, billiard balls, and mechanical devices, were inspired by memories from the father’s biography. Since actual photos from his childhood were unavailable, AI generative tools recreated these items as “old photos” based on the father's recollections, annotated with contextual stories (Fig. 4). These props enable players to communicate with YELL and uncover clues about his psychological struggles.

The final scene removes YELL entirely, replacing him with a real family documentary, marking the game’s emotional climax (Fig. 6). A vintage television screen plays a 1998 home video of the artist’s family at a children’s park, capturing moments of shared joy. Through montage editing, the artist re-contextualizes these interactions, bridging the divide between fiction and reality. This conclusion subverts the game’s symbolic framework, revealing that beneath the constructed layers of narrative and AI simulation lies an authentic familial bond. The real, in this moment, breaks through the symbolic, leaving players with a profound emotional resolution.


\begin{figure}[h]
\includegraphics[width=3.35in]{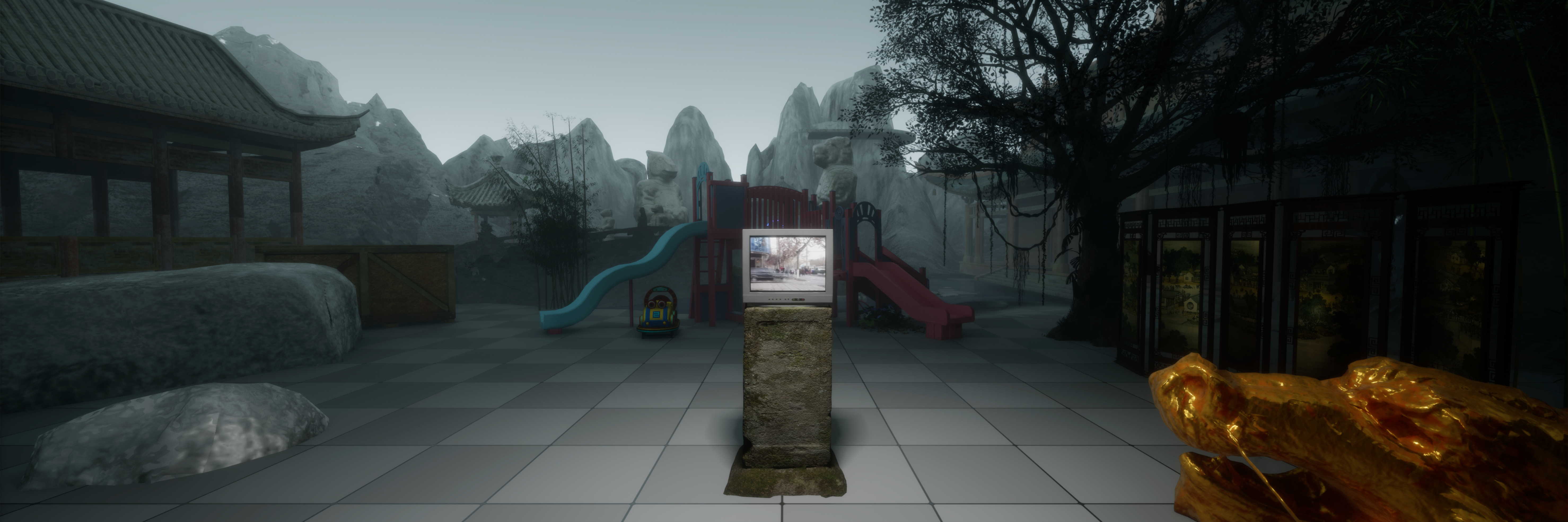}
\caption{Still frame from the final scene.} 
\end{figure}

\begin{figure}[h]
\includegraphics[width=3.31in]{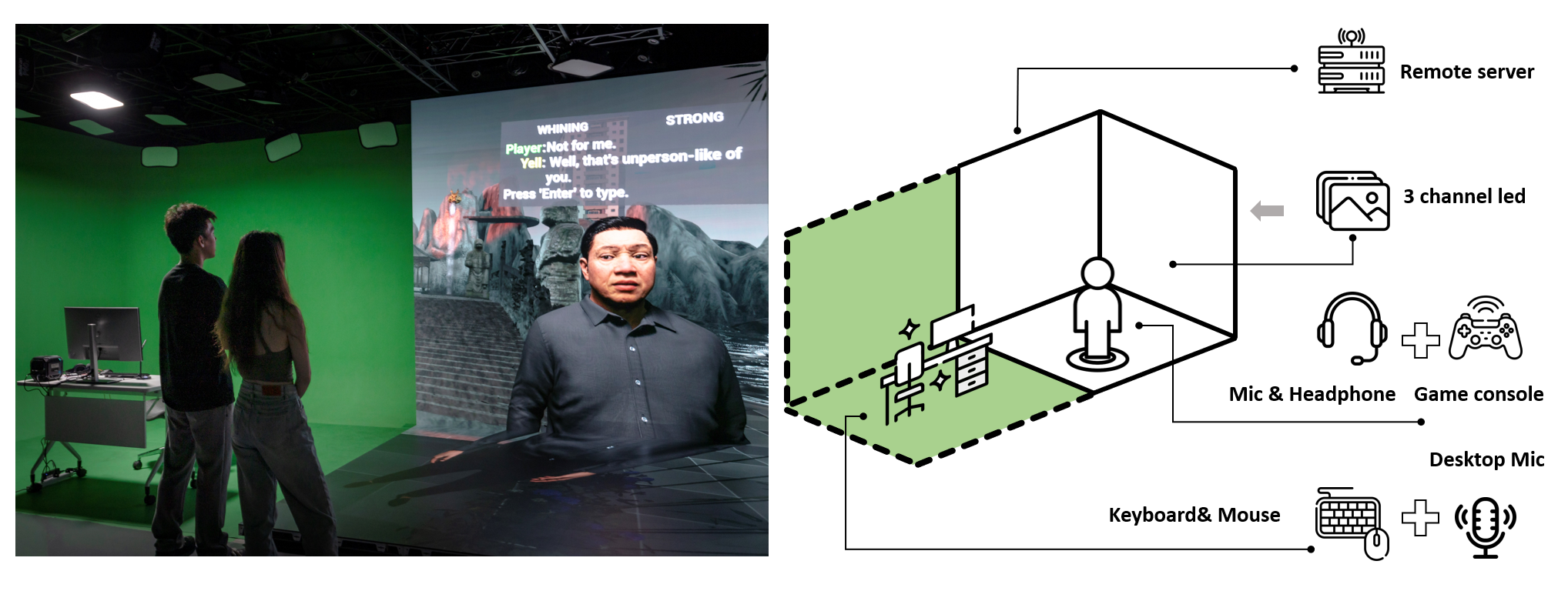}
\caption{The final display of the interactive experience in the Cave system.} 
\end{figure}

\subsection{Technical implementation}
To enhance the blending of reality and virtual elements, \emph {The Dream Within Huang Long Cave} was presented in a hybrid environment, combining a CAVE system for an immersive experience. This arrangement involved two game programs running concurrently (Fig. 7). The first setup was hosted in the CAVE system, where the game operated on a remote server. Players controlled the first-person perspective using a game controller and wireless microphone. With Inworld AI's multilingual feature, YELL responded to real-time speech recognition in English, Chinese, Korean, and Japanese. In Unreal Engine, we implemented a multi-perspective, multi-window rendering scheme to fully utilize the CAVE’s spatial and visual dimensions, enhancing both field of view and immersion.

Simultaneously, another version of the game ran locally on a PC, positioned centrally within the green screen space adjacent to the CAVE. This configuration allowed an additional player to engage with the game concurrently. In the local setup, the player interacted using a mouse and keyboard, with the option to input text either by typing or via voice commands. Although both platforms operated independently, their juxtaposition symbolized the dialogic relationship presented within the game, offering distinct yet interconnected player experiences.  

\section{Results}

As an outcome of this psychoanalytic artistic practice, the project resulted in two distinct works: an interactive narrative experience and a Machinima film. The first work, \emph{The Dream Within Huang Long Cave}, is an interactive and immersive experience presented in the format of a video game. Through AI-driven dialogue, players engage with YELL exploring his fragmented memories and unconscious struggles. This real-time interaction allows audiences to uncover YELL’s unresolved emotional conflicts while reflecting on their own familial experiences.

The second work, a Machinima film derived from the gameplay, serves a different yet complementary function \footnote{\url{https://vimeo.com/996047826/6203b02113}}. Featuring the artist as the director, son, and player, the film captures a recorded playthrough where the artist engages in dialogue with the AI-father. Functioning both as a playthrough and an experimental video artwork, the film encapsulates a cross-temporal conversation, reflecting the artist’s personal reconciliation process. Unlike the interactive game, where players navigate the experience themselves, the Machinima film curates a structured narrative, highlighting the evolving emotional dynamics between the artist and the digital father.

Both works were exhibited in different contexts, reaching varied audiences (Fig. 8a). The interactive narrative was featured in a solo exhibition at Three Shadows Photography Art Centre in Beijing, where it was showcased alongside six related works exploring themes of family memory and psychoanalysis \footnote{\url{https://vimeo.com/1013824598}}. Additionally, the project was included in the 10th Jimei × Arles International Photo Festival, as part of a group exhibition focusing on personal and familial recollection. The Machinima film was selected for the HiShorts! Film Festival’s experimental film track, where it was screened in a theater, engaging a broader audience beyond interactive media.

The project received diverse responses from curators, artists, and general audiences. A curator remarked that the dreamlike journey through Huang Long Cave evoked Lacan’s concept of mi-dire (half-saying), where real emotion emerges through the gaps between what is said and what remains unsaid. A senior artist described the work as evoking a deep sense of pain, identifying the portrayal of familial distance as a universal emotional experience. In addition to formal critiques, the exhibition also accumulated anonymous audience feedback, revealing moments of personal reflection, familial contemplation, and spontaneous emotional reactions. This suggests that the interactive structure successfully prompted audiences to engage with personal and collective memories, affirming the project’s ability to challenge rigid paternal structures while fostering emotional reflection.

\begin{figure}[h]
\includegraphics[width=3.31in]{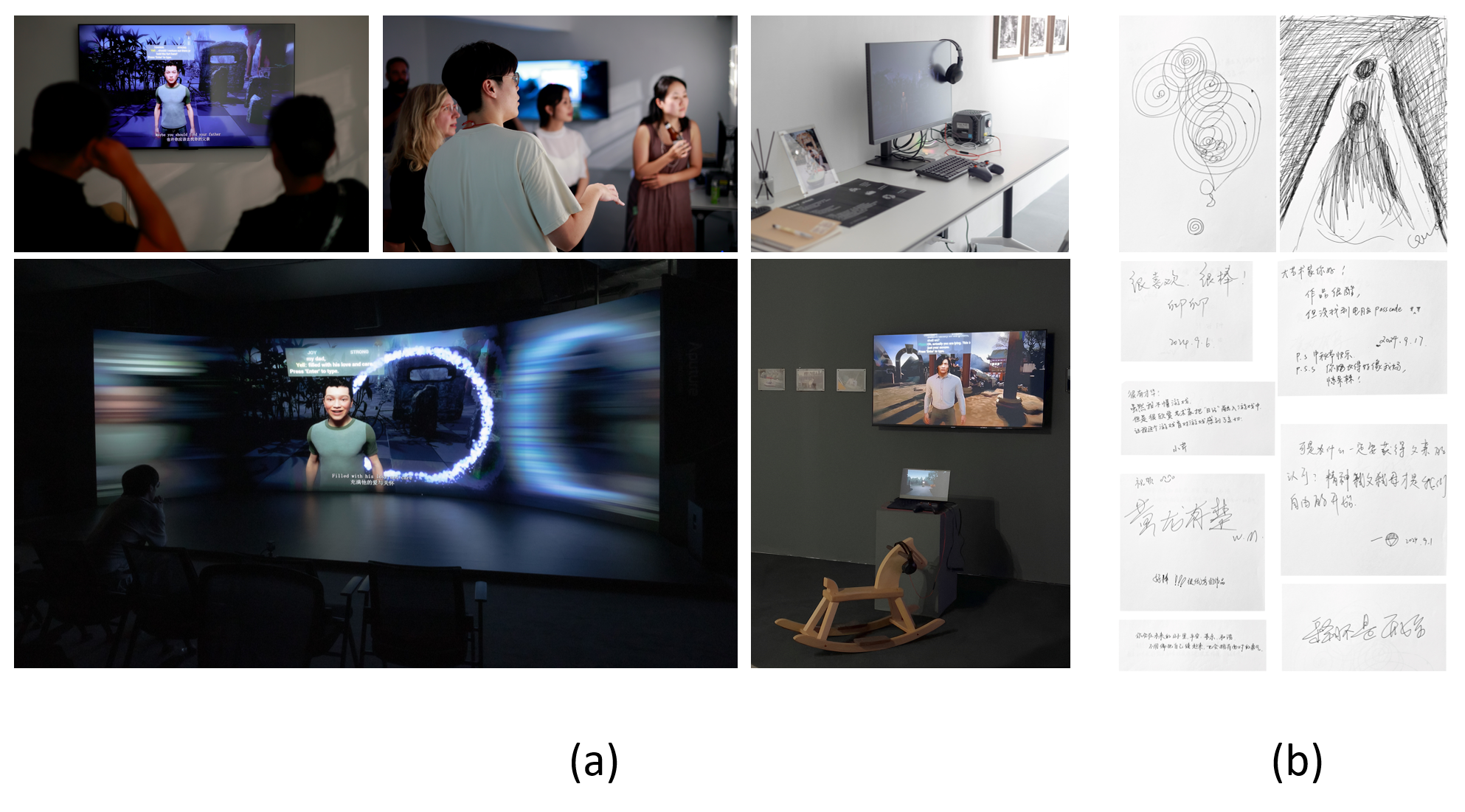}
\caption{(a) Photos from exhibitions and screenings. (b) Anonymous audience comments.} 
\end{figure}

\section{Discussion}
One of the key discoveries in this project was how the interactive storytelling process became a bridge for father-son communication. Initially, this project was conceptualized as a way to examine paternal authority, yet it unexpectedly provided a symbolic mourning process for the artist’s father. When the father lost his own father, it left a psychological wound that remained unresolved for decades. Through participation in this project, he engaged with the digital representation of a father figure, allowing him to reflect on his own role in the cycle of fatherhood. From a psychoanalytic perspective, this project demonstrates that fatherhood is not a fixed role but a symbolic position that shifts over time. In traditional families, the grandfather is the previous Big Other, the figure who structures the family's symbolic order. However, when this first Big Other is lost, the family must reconstruct its symbolic stability. In this case, the father, once positioned as a “son,” was forced to assume the role of the Big Other. This transition was emotionally complex, as he found himself embodying the very authority he once struggled against. The project’s interactive structure allowed him to revisit his father’s presence in a symbolic rather than purely memorializing manner, demonstrating that familial roles are constantly restructured through the processes of loss, inheritance, and symbolic reconstruction.

On the other hand, audience responses revealed how the project resonated beyond traditional father-son relationships. One particularly insightful comment came from a female participant raised in a single-parent household. She noted that, unlike YELL, she had no personal memories of a father figure, making the game’s premise initially feel distant. However, as she engaged further, she reflected on how her mother had, in many ways, fulfilled the symbolic role of the father in her life. This realization led her to consider how authority, discipline, and family dynamics are structured even in non-traditional family arrangements. This feedback reinforced a core idea of this project: fatherhood is not merely a biological reality but a symbolic function. Even in families where the biological father is absent, the symbolic structure often remains intact through other figures—mothers, grandparents, or guardians—who occupy the role of the Name-of-the-Father. This reflection suggests that while psychoanalysis historically assumes the presence of the father in structuring subjectivity, contemporary families reveal a broader spectrum of figures who enact the function of paternal authority.

Finally, this project also raises questions about the ethical implications of AI in psychoanalytic art practice. A recurring question among participants was whether an AI-driven entity, if provided with enough personal data, could truly replicate a person’s consciousness. This aligns with broader concerns in AI research about digital immortality—the notion that a person’s identity can be preserved indefinitely through machine learning models. However, from a psychoanalytic perspective, such a proposition raises ethical concerns. Lacan’s ethics emphasize that human subjectivity is defined by its “lack,” meaning that no entity—biological or artificial—can ever fully encapsulate a subject’s unconscious drives and desires. If an AI were to completely fill this lack, it would risk collapsing the fundamental gap that defines human subjectivity. Instead, AI should be understood not as a substitute for the lost other, but as a symbolic representation that mediates between memory and presence. This suggests that rather than aiming for perfect digital preservation, AI should serve as a means of facilitating real emotional connection with absence.

\section{Conclusion}
This paper proposes \emph {The Dream Within Huang Long Cave,} an interactive narrative experience, as an artistic response to Lacan’s proposition of “the non-existence of the Big Other.” The study investigated the shifting symbolic role of fatherhood, the evolving structure of contemporary families, and the ways in which AI-driven storytelling can challenge and reframe traditional paternal authority. By deconstructing the illusions of the "Name-of-the-Father" and the "Big Other," the project offered a dynamic exploration of family relationships, demonstrating that paternal authority is not a fixed role but a symbolic position that transforms over time.

This research also introduced the “Analytic-Critical Method”, combining psychoanalysis, practice-based art, and AI-driven technology to construct a digital paternal figure. By using AI language models as mediators of the symbolic Big Other, the project redefined interactive psychoanalytic storytelling as a means of emotional exploration. Audience responses underscored the relevance of these themes beyond traditional father-son dynamics, highlighting how paternal authority manifests across diverse family structures.

Beyond narrative and artistic contributions, this study raised ethical concerns regarding AI, digital subjectivity, and memory preservation.  Instead of striving for perfect digital replication, this project suggests that AI should serve as a mediator between memory and presence, fostering real emotional connections rather than replacing them. Ultimately, this research highlights the potential of psychoanalytic artistic practice to bridge memory, technology, and emotional experience, offering new perspectives on interactive storytelling, AI ethics, and family relationships.




\bibliographystyle{isea}
\bibliography{isea}

\end{document}